\input{aipcheck}

\documentclass[
    ,final            
  ]
  {aipproc}

\layoutstyle{6x9}

\usepackage[numbered,framed]{mcode}

\begin{document}

\title{A new gravitational N-body simulation algorithm for
investigation of cosmological chaotic advection}

\classification{98.80.-k,98.65.At.Cw.Dx,95.35.+d,95.75.Pq}
 
 
 


\keywords      { Gravitational N-body simulation, Large-scale structures, Alternative cosmological models, GPU computing}

\author{Diego H. Stalder}{
  address={Lab for Computing and Applied Math (LAC) - National Institute for Space Research (INPE),\\ S\~ao Jos\'e dos Campos, SP, Brazil}
}

\author{Reinaldo R. Rosa}{
  address={Lab for Computing and Applied Math (LAC) - National Institute for Space Research (INPE),\\ S\~ao Jos\'e dos Campos, SP, Brazil}
}

\author{Jos\'e R. da Silva Junior}{
  address={Computing Institute (IC) - Fluminense Federal University  (UFF),\\ Niteroi, RJ, Brazil}  
}

\author{Esteban Clua}{
  address={Computing Institute (IC) - Fluminense Federal University  (UFF),\\ Niteroi, RJ, Brazil}
}
\author{Renata S. R. Ruiz}{
  address={Lab for Computing and Applied Math (LAC) - National Institute for Space Research (INPE),\\ S\~ao Jos\'e dos Campos, SP, Brazil}
}
\author{Haroldo F. Campos Velho}{
  address={Lab for Computing and Applied Math (LAC) - National Institute for Space Research (INPE),\\ S\~ao Jos\'e dos Campos, SP, Brazil}
}
\author{Fernando M. Ramos}{
  address={Lab for Computing and Applied Math (LAC) - National Institute for Space Research (INPE),\\ S\~ao Jos\'e dos Campos, SP, Brazil}
}
\author{Amarísio da Silva Araújo}{
  address={Lab for Computing and Applied Math (LAC) - National Institute for Space Research (INPE),\\ S\~ao Jos\'e dos Campos, SP, Brazil}
}
\author{Vitor Conrado F. Gomes }{
  address={Lab for Computing and Applied Math (LAC) - National Institute for Space Research (INPE),\\ S\~ao Jos\'e dos Campos, SP, Brazil}
}

\begin{abstract}
Recently alternative approaches in cosmology seeks to explain the nature of dark matter as a direct result of the non-linear spacetime curvature due to different types of deformation potentials. In this context, a key test for this hypothesis is to examine the effects of deformation on the evolution of large scales structures. An important requirement for the fine analysis of this pure gravitational signature (without dark matter elements) is to characterize the position of a galaxy during its trajectory to the gravitational collapse of super clusters at low redshifts. In this context, each element in an gravitational N-body simulation behaves as a tracer of collapse governed by the process known as \textit{chaotic advection} (or \textit{lagrangian turbulence}).  In order to develop a detailed study of this new approach we develop the COsmic LAgrangian TUrbulence Simulator (COLATUS) to perform gravitational N-body simulations based on Compute Unified Device Architecture (CUDA)  for  graphics processing units (GPUs). In this paper we report the first robust results obtained from COLATUS.
 \end{abstract}

\maketitle

\section{I. Introduction}
The gravitational N-body simulations have become a powerful tool for testing the theories of structure formation in astrophysical and cosmological systems \cite{Bertschinger:1996}. In particular, it has been shown that the statistical characterization of dark matter distribution is an important ingredient in the investigation of large-scale structure formation in the Hubble volume simulated from the GADGET-VC algorithm \cite{Carettaetall:2008}. 

Recently, an established statistical method was used to demonstrate the importance of considering chaotic advection (or \textit{Lagrange Turbulence}) \cite{Aref:2002} in combination with gravitational instabilities in the $\Lambda$ - CDM simulations performed from the Virgo Consortium (VC) \cite{Rosaetall:2009}. However, the GADGET-VC algorithm does not allow in a straightforward approach the computation of the kinematics of a single particle, requirement which is necessary for the investigation of the chaotic advection in alternative cosmological models considering only barionic matter \cite{Rosa:2010}. This limitation appears because the interaction forces are computed by the TreePM scheme  \cite{Springel:2005}. The COsmic LAgrangian TUrbulence Simulator (COLATUS) is a new algorithm to perform gravitational N-body simulations allowing the computation of the velocity of a single particle at every time-step and then the evaluation of its energy power spectrum. To achieve its objective COLATUS compute the gravitational forces by using a direct summation scheme(the simplest PP scheme). COLATUS is implemented in a Compute Unified Device Architecture (CUDA) by using the Nvidia graphics processing units (GPUs) to reduce the simulation runtime. In the present work we show the preliminary simulations including up to $32^3$ particles using 1536 cores of NVIDIA GTX680. 

 \section{II. GPU/CUDA COSMOLOGICAL N-BODY SIMULATION}
 
 For N-bodies under the influence of physical gravitational forces in a comoving coordinates($\overrightarrow{x_{k}}=\overrightarrow{r_{k}}/a$) the motion equations take the form:
\begin{eqnarray}\label{nbodyed11} 
\frac{d\overrightarrow{x_{k}}}{dt}=\overrightarrow{v_{k}},\quad
\frac{d\overrightarrow{v_{k}}}{dt}+2\frac{\dot{a}}{a}\overrightarrow{v_{k}}=-a^{-3}\nabla_{x} \phi =a^{-3}G\sum^{N}_{k\neq j}\frac{m_{j}(\overrightarrow{x_{k}}-\overrightarrow{x_{j})}}{(\mid \overrightarrow{x_{k}}-\overrightarrow{x_{j}}\mid^{2}+\epsilon^{2})^{3/2}},\quad\mbox{for}\quad k=1,\dots ,N.
\end{eqnarray}
where $\epsilon$ is the softening factor(collisionless assumption), $a$ is the scale factor which is taken to be $1$ at the present time and it changes ruled by the Friedmann equation:

\begin{eqnarray}\label{nbodyed}
\frac{\dot{a}}{a}=H(t)=H_{0}\sqrt{\Omega_{R}\,a^{-4}+\Omega_{M}\,a^{-3}+\Omega_{K}\,a^{-2}+\Omega_{\Lambda}}\quad\quad \end{eqnarray}
where $\Omega_{R}$ is the radiation density, $\Omega_{M}$ is the matter (dark plus baryonic) density,  $\Omega_{K}$ is spatial curvature density, $\Omega_{\Lambda}$  is the cosmological constant or vacuum density today and $H_{0}$ is the Hubble parameter for $z=0$. These equations are valid
for non-relativistic matter $(v << c,\Phi << c^{2})$ at scales that are much smaller than the Hubble radius $(L << c/H_{0})$ \cite{Mccrea:1934}. The equation of motions can be integrated using
the difference equations\cite{HockneyandEastwood:1981}.
 \begin{eqnarray}
 \overrightarrow{x}^{n+1/2}&=& \overrightarrow{x}^{n}+\frac{1}{2}\tau  \overrightarrow{v^{n}}	\\
 \overrightarrow{v}^{n+1}&=& \overrightarrow{v}^{n}\frac{1-H(t)\tau}{1+H(t)\tau}+\frac{\nabla_{x} \phi(\overrightarrow{x}^{n+1/2})\tau}{a^{3}(1+H(t)\tau)}	\\
 \overrightarrow{x}^{n+1}&=& \overrightarrow{x}^{n+1/2}+\frac{1}{2}\tau  \overrightarrow{v}^{n+1}.	
 \end{eqnarray}
 However, these equations are not symplectic, but by making a suitable canonical transformation, one can derive an integrator that is symplectic \cite{Quinnetall:1997,Chambers:1998}. 

The Lagrangian for the particle motion in the comoving frame is:
 \begin{eqnarray}
\mathcal{L}=\frac{1}{2}m\,a^{2}\,\overrightarrow{v}^{2}-m\frac{\phi\prime(\overrightarrow{x})}{a}\quad\quad
\phi\prime=a\phi+\frac{1}{2}a^{2}\dot{\dot{a}}
 \end{eqnarray}
Switching to the Hamiltonian formalism, the
momentum canonical to $\overrightarrow{x}$ is $\overrightarrow{p}=m\,a^{2} \overrightarrow{v}$ and the Hamiltonian is:
 \begin{eqnarray}
H=\frac{\overrightarrow{p}^{2}}{2\,m\,a^{2}}+m\frac{\phi\prime(\overrightarrow{x})}{a}
 \end{eqnarray}
then 
 \begin{eqnarray}
 \frac{d\overrightarrow{x}}{dt}=\frac{dH}{d\overrightarrow{p}}=\frac{\overrightarrow{p}^{2}}{2\,m\,a^{2}}\quad\quad
\frac{d\overrightarrow{p}}{dt}=-\frac{dH}{d\overrightarrow{x}}=-m\frac{\phi\prime(\overrightarrow{x})}{a}.
 \end{eqnarray}
Although this Hamiltonian is time-dependent, it
is separable, so the ``drift'' and ``kick'' operators
are easily derived as:
   \begin{eqnarray}
 D(\tau)=\overrightarrow{x}^{t+\tau}=\overrightarrow{x}^{t}+\overrightarrow{p}\int_{t}^{t+\tau}\frac{dt}{m\,a^{2}}\quad\quad
K(\tau)=\overrightarrow{p}^{t+\tau}=\overrightarrow{p}^{t}-m\phi\prime(\overrightarrow{x})\int_{t}^{t+\tau}\frac{dt}{a}
 \end{eqnarray}
 where $\tau$ is the step size.

Note that, the gravitational force in the Newtonian limit falls as $1/r^{2}$, hence it is a long range force and we cannot ignore force due to distant particles. This makes calculation of forces
the most time consuming task in N-Body simulations. As a result, a lot of attention has been focused on this aspect and many algorithms and optimizing schemes have been developed, we refer the reader to \cite{Bertschinger:1996,Bagla:2005} for a detailed review. These schemes have been evolved to address this problem and replace direct summation by methods that are less time consuming. However 
the direct summation implemented in GPU/CUDA approach provide at least three features that help it achieve high efficiency \cite{Nyland:2007}: (i) Straightforward parallelism with sequential memory access patterns; (ii) Data reuse that keeps the arithmetic units busy; and (iii) The fully pipelined arithmetic, including complex operations such as inverse square root, that are much faster clock-for-clock on a GPU than on a CPU.

The standard model that explains the large scale structure formation (e.g. galaxies, clusters of galaxies) assume that they were formed from the amplification of the small initial fluctuations via gravitational instability \cite{Jeans:1902}. These models attempt to reduce cosmology to an initial value problem \cite{Bertschinger:1996}. Then, given the initial conditions(universe at high redshift) and a cosmological model, the goal is to compute the evolution of structure until $z=0$. The initial conditions are a density-velocity field represented by a set of particles which must be set-up in compatibility with of the observed primordial universe on cosmic microwave background radiation \cite{Zeldovich:1970,Guth:1980,Jeong:2010}. We generate the initial conditions by imposing perturbations on an initially uniform state represented by a ``glass'' distribution of particles generated by the method of White (1996). Using the algorithm described by \cite{Efstathiouetall:1985}, which is based on the Zeldovich approximation \cite{Zeldovich:1970}, a Gaussian random field is set up by perturbing the positions of the particles and assigning them velocities according to growing mode linear theory solutions \cite{Jenkinsetall:1998}. 

 The cosmological models specifies the universe composition(matter, radiation, spacetime curvature) and attempt to describes its evolution and interaction, they usually include some exotic ingredients as the dark matter and dark energy to explain anomalous galactic rotation curves and the accelerated expansion, respectively. Assuming that the gravitation is the dominant force (e.g.\cite{Bagla:2005}) we compute the structure evolution by integrating the motion equations and computing the N-body interaction forces in every time-step. The motion equations are in comoving spatial coordinates to consider the universe expansion and boundary conditions are periodic to be consistent with a homogeneous elements distribution \cite{Bertschinger:1996,Bagla:2005,Jenkinsetall:1998}. The solutions of the motion equations would be exact if we were able to simulate the
motions of all individual bodies. Unfortunately this can not be achieved
since the systems of interest (cosmological large scale structures) contain of the order of $10^{20}$ stars.
We approximate collection of a very large number of stars in the universe by one galaxy
in an N-Body simulation. Therefore the particles in an N-Body simulation must interact in a
purely collisionless manner \cite{BaglaandPadmanabhan:1997,Moetall:2010}. The mass of a galaxy (typically of the order of 0.1 to $1\times 10^{12}\,M_{\odot}$ ) is adjusted in its normalized form (0.1-1) in the initial condition hypercube.

Based on the formalism described above we have implemented the algorithm COLATUS that is described in a simplified form in Appendix A.

\section{III. Preliminary Results}
 
 The first main objective to be reached with the COLATUS is to show that simulations with the simplest type of particle-particle (PP) gravitational interaction, containing typical densities of barionic content only, one can get collapsed patterns consistent with gravitational simulations that include dark matter.

\begin{figure}[h]
  \includegraphics[height=.28\textheight]{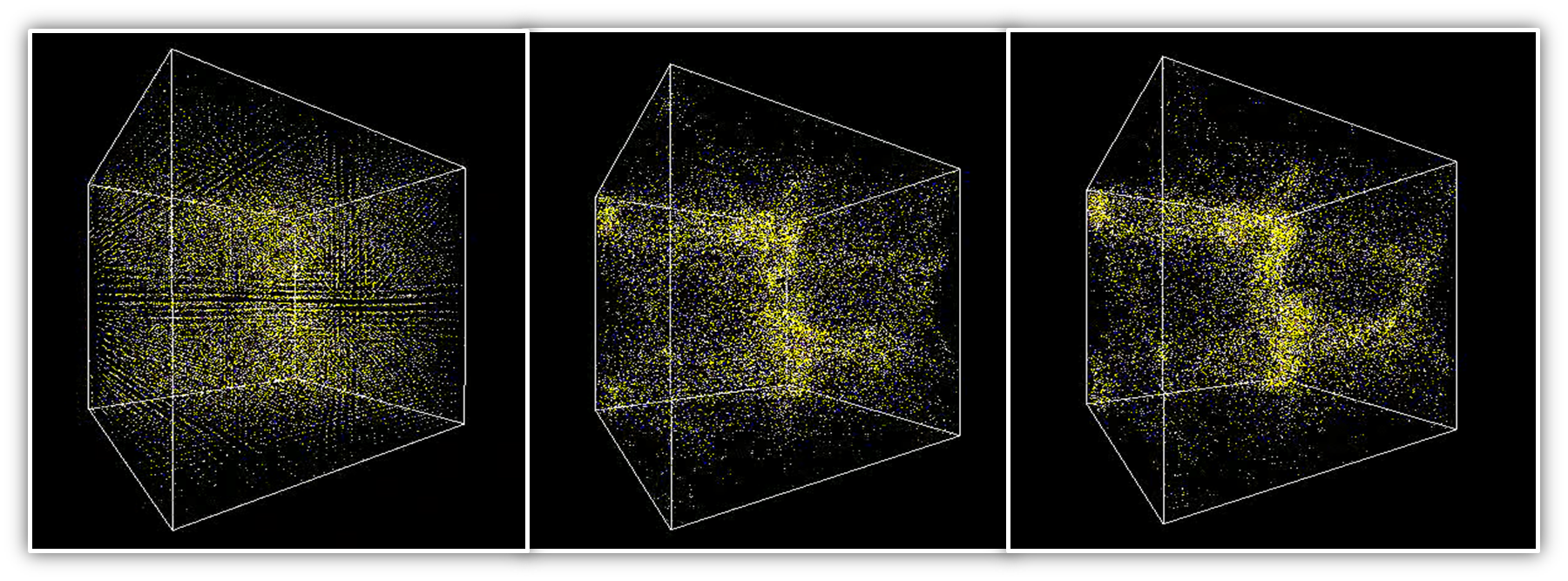}
  \caption{Three snapshots from a COLATUS simulation with $N=32^{3}$, Box size $L= 100 Mpc$, IC= Eisenstein and Hu Spectrum ($z=50$).}
\end{figure}

Figure 1 depicts three snapshots of the  large-scale structure formation for a cosmological box containing $32^{3}$ galaxies in the critical density distributed in a volume of $(100Mpc)^{3}$ which evolves over $51$ units of redshift. The spectrum for the initial distribution of particles follows  the Eisenstein-Hu law \cite{EisensteinandHu:1997}. 
For $z=0$ we have computed the relation between the number of filaments and the number of voids, 
$\frac{N_{f}}{N_{v}}=\frac{3}{5}$, and  the averages for the filament lengths $L_{f}=54Mpc$, filament thickness $L_{ft}=13.3Mpc$ and  void lenghts $L_{V}=39,33Mpc$, which are
consistent with those obtained by simulations based on more sophisticated supercomputers (see for example \cite{Springel:2005} Virgo and Millenium).

For the next simulations we intend to begin the study of the deformation potential influence on the large scale structures patterns (filament thickness, filament and void lenghts)  obtained for z = 0. In order to implement this study we will modify the equation (\ref{nbodyed11}) by introducing a deformation potential $-\frac{\psi(\overrightarrow{x_{k}}-\overrightarrow{x_{\psi})}}{(\mid \overrightarrow{x_{k}}-\overrightarrow{x_{\psi}}\mid^{2}+\epsilon^{2})^{3/2}}$ as follow:
\begin{eqnarray}\label{nbodyed} 
\nabla_{x} \phi_{\psi} =-G\sum^{N}_{k\neq j}\frac{m_{j}(\overrightarrow{x_{k}}-\overrightarrow{x_{j})}}{(\mid \overrightarrow{x_{k}}-\overrightarrow{x_{j}}\mid^{2}+\epsilon^{2})^{3/2}}-C\frac{\psi(\overrightarrow{x_{k}}-\overrightarrow{x_{\psi})}}{(\mid \overrightarrow{x_{k}}-\overrightarrow{x_{\psi}}\mid^{2}+\epsilon^{2})^{3/2}}.
\end{eqnarray}
where $C$ is a constant and  $x_{\psi}$ is the position where the potential has it max value. Note
that potential force fall as $1/r^{2}$, then it acts like a mass particle located in a fixed position.

 \section{IV. Concluding Remarks} 
 
 We show here a new way to apply simulations of gravitational N-body using CUDA / GPU to test cosmological models that attempt to explain dark matter from effects due only to the curvature of spacetime. Preliminary results indicate that the algorithm is able to simulate the generation of large-scale structures whose statistical properties are consistent with simulations that use of more sophisticated numerical schemes for supercomputers based on CPUs.

An important fact in this new algorithm is that it allows to directly follow the trajectory of individual galaxies during the evolution of the universe, thus allowing to test, with high precision, the effects of possible global and local deformations of the Lagrangian particle description in spacetime. In this context, the statistical methods for structural  fine analysis should be those commonly used to characterize the process of chaotic advection which tracers moving in turbulent fluids are subject. In our cosmological approach, galaxy clusters are formed only by baryonic matter that also feel the effect of deformation that will intensify the effect of gravitational instabilities inherent in the system.

\section{Appendix A} 

The COLATUS algorithm implement the direct summation scheme by using a code written in CUDA.
 On the GPUs each thread compute the force, acceleration, position and velocity of each particle. 
The sequence of the algorithm is as follows:

\begin{enumerate}

\item First, the CPU reads the initial conditions from a datafile and copy the initial positions and velocities to the global memory;

\item On the CPU, the main program allocate the Memory on the GPU;

\item On the CPU, the main program copy the initial conditions to the GPU;

\item On the CPU, the main program initialize the threads on the GPU each of them will perform the calculations;

\item On the GPU, the forces are calculated by the threads.

\item On the GPU, we integrate the motion equation to actualize the particles velocities and positions. We are testing the viability of make the numerical integration on the CPU, then we will use the GPU only for calculation of the iteration forces (e.g. \cite{Hamadaetall:2010})).

\item The process is repeated and concurrently other library shows a particle motion in the box.

\item At every time-step the GPU copy to the CPU the particle position and velocity.

\end{enumerate}

Below is an excerpt from the main part of the code in CUDA which is responsible for calculating the particle-particle gravitational interaction.

\begin{lstlisting}
__device__ float4 computeDistance(float4 pos1, float4 pos2, 
float min_distance) {
...
}
__global__ void force_calculation(float4* position,float4* velocity,
        float4* accell, int numbodies, float g,
     bool useCollision, float min_dist, float dt, float3 box_dim){
          int bx   = blockIdx.x;  int tx   = threadIdx.x;
          int dimX = blockDim.x;  int idx  = bx * dimX + tx;
          float4 acc = make_float4(0,0,0,0);
          float4 netForce = make_float4(0,0,0,0);
          float4 pos = position[idx];   float4 vel = velocity[idx];
for (int i=0; i<numbodies; i++) { 
      if (idx != i) {
  float4 directionDist = computeDistance(position[idx], position[i]
              ,min_dist); float mass2 = position[i].w;
             Force.x += g*mass2 * directionDist.x * directionDist.w;
             Force.y += g*mass2 * directionDist.y * directionDist.w;
             Force.z += g*mass2 * directionDist.z * directionDist.w;
                } }
//Numerical Integration
..
}
\end{lstlisting}

\begin{theacknowledgments}
  The authors thank the scientific Brazilian support agencies: CAPES, CNPq and FAPESP.

\end{theacknowledgments}



\bibliographystyle{aipproc}   


\end{document}